\documentclass{iopconfser}

\usepackage{amssymb,amsmath,mathrsfs,amsthm}
\usepackage{graphicx,wrapfig}
\usepackage{tikz}
\usepackage{hyperref}
\hypersetup{
	breaklinks=true,  
	colorlinks=true,   
}
\usepackage{doi}
\usepackage{xurl}
\usepackage{breakurl}
\usepackage{orcidlink}

\newcommand{\HH}{\mathcal{H}}
\newcommand{\MM}{\mathcal{M}}
\newcommand{\KK}{\mathcal{K}}

\newcommand{\RR}{\mathbb{R}}

\newcommand{\nml}{\textup{n}}
\newcommand{\diff}{\textup{d}}

\newcommand{\Rcal}{\mathcal{R}}
\newcommand{\Scal}{\mathcal{S}}
\DeclareMathOperator{\Bor}{Bor}
\DeclareMathOperator{\Aut}{Bor}
\DeclareMathOperator{\Ad}{Ad}
\DeclareMathOperator{\id}{id}

\newcommand{\ol}[1]{\overline{#1}}

\newcommand{\Af}{\mathscr{A}}

\newtheorem{theorem}{Theorem}

\begin{document}

\title{Quantum reference frames for spacetime symmetries and large gauge transformations}

\author{Daan W.~Janssen$^{1}$ \orcidlink{0000-0001-7809-5044}\\\textnormal{Based on joint works with Christopher J. Fewster, Leon D. Loveridge, Kasia Rejzner and James Waldron}}

\affil{$^1$Department of Mathematics, University of York, United Kingdom}

\email{daan.janssen@york.ac.uk}
\begin{abstract}
Symmetries are a central concept in our understanding of physics. In quantum theories, a quantum reference frame (QRF) can be used to distinguish between observables related by a symmetry. The framework of operational QRFs provides a means to describe observables in terms of their relation to a reference quantum system. We discuss a number of applications of QRFs in the context of quantum field theory on curved spacetimes: 1) A type reduction result for algebras arising from QFTs and QRFs with good thermal properties. 2) Quantisation of boundary electric fluxes and gluing procedures for quantum electromagnetism on spacetimes with boundaries.
\end{abstract}
\section{Why (quantum) reference frames?}
Throughout the history of physics, symmetries have played a central role. They often serve as a tool to simplify the analysis of a model and yield conservation laws \cite{noetherInvarianteVariationsprobleme1918}. More fundamentally, symmetries offer a guiding principle in the construction of physical theories, not least for general relativity \cite{einsteinFeldgleichungenGravitation1915} and gauge theories \cite{yangConservationIsotopicSpin1954}. In many a setting, physical observables of interest are those that are invariant under a symmetry. In special relativity for instance \cite{einsteinZurElektrodynamikBewegter1905}, one may take physical quantities to be those that are Poincar\'e invariant.\footnote{Due to P\&T symmetry violations in particle physics (see e.g.~\cite{aaijObservationChargeParity2025}), such an invariance requirement can reasonably be restricted to (time-)orientation preserving transformations.} In practice, observables of interest are defined relative to some inertial observer, measurement apparatus or some other reference object, fixing a preferred inertial reference frame. This is compatible with the invariance requirement as long as, when the reference object taken as part of the system, these relative quantities are again Poincar\'e invariant.

We often assume that the objects used to define a reference frame (an observer, a measurement apparatus, a clock, a rod etc.) are described by classical physics. Yet in recent years, considerable effort has gone into describing reference frames in quantum physics. Since early work dating back to the 1980's \cite{aharonovQuantumFramesReference1984}, numerous notions of quantum reference frames (QRFs) have been introduced \cite{bartlettReferenceFramesSuperselection2007,loveridgeSymmetryReferenceFrames2018a,giacominiQuantumMechanicsCovariance2019,vanrietveldeChangePerspectiveSwitching2020}.  Here we discuss some recent applications of this concept in the context of quantum field theory (QFT) on curved spacetimes, with a focus on algebraic features of these theories. In one application, we consider a QFT together with an external QRF, and see that the mathematical structure of invariant observables of this joint QFT-QRF system can differ significantly from QFT observables defined relative to a classical reference frame \cite{fewsterQuantumReferenceFrames2025}. In some examples, we see this opens up new possibilities to defining entropy for quantum fields. In a second application, we see that QRFs provide a natural way to describe edge modes of gauge QFTs on spacetimes with boundaries and corners. In the case of electromagnetism, the inclusion of such edge mode QRFs into the algebra of observables, results in a quantisation of the electric flux through the spacetime corner \cite{fewsterSemilocalObservablesEdge2025}.

To describe quantum fields on curved spacetime, we use the formalism of \emph{algebraic quantum field theory} (AQFT) as our starting point \cite{brunettiAdvancesAlgebraicQuantum2015}. This is in line with our focus on algebraic properties of QFT-QRF models. To describe the frames, we make use of the \emph{operational QRF} formalism \cite{loveridgeSymmetryReferenceFrames2018a,caretteOperationalQuantumReference2025,glowackiQuantumReferenceFrames2024a}. Similarly to AQFT, this is the approach to QRFs most naturally formulated using the language of operator algebras. We briefly describe this operational formalism below.

\section{Operational QRFs}
The rough idea of an operational QRF, is that it is an object that assigns to each quantum state a probability distribution over classical reference frames. For the Poincar\'e group, a sensible notion of classical reference frame is simply a choice of inertial coordinates on Minkowski space. Similarly, for a group of time translations on some static spacetime, a classical reference frame simply amounts to a choice of global time coordinate. On a more abstract level, we take classical reference frames to form a homogeneous $G$-space for the symmetry group $G$. For simplicity of our discussion, we shall restrict our attention to cases where we may identify this space with $G$. For technical reasons, we shall furthermore assume that $G$ is a locally compact topological group. 

Following \cite{caretteOperationalQuantumReference2025}, we formulate the notion of QRFs in terms of bounded operators on a Hilbert space. We assume that the observables of the reference system defining the QRF are described by some bounded operators on a Hilbert space $\HH_{\Rcal}$. The group $G$ acts on $B(\HH_{\Rcal})$ via the adjoint action of some appropriately continuous unitary representation $U_{\Rcal}:G\mapsto B(\HH_{\Rcal})$. We now say that a QRF for $G$ is a normalised positive operator valued measure $E:\Bor(G)\to B(\HH_{\Rcal})$ such that for each $g\in G$ and $S\in \Bor(G)$ a Borel subset, we have a covariance relation $	U_{\Rcal} E(S) U_{\Rcal}^*=E(g.S)$. Indeed, if we think of elements of $G$ (as a homogeneous $G$-space) as classical reference frames, a normal state $\omega_{\Rcal}$ on $B(\HH_{\Rcal})$ will define a probability distribution $\omega_{\Rcal}\circ E$ on $G$. We refer to the triple $(\HH_{\Rcal},U_\Rcal,E_\Rcal)$ as a QRF for $G$.

The main purpose of an operational QRF is to ensure existence of non-trivial $G$ invariant observables built from a $G$-covariant quantum system and the QRF. 
To be more concrete, we consider a quantum system (possibly a QFT) whose observables form a von Neumann algebra $\MM_{\Scal}\subset B(\HH_{\Scal})$ on some Hilbert space $\HH_{\Scal}$ \cite{Takesaki2001}. 
We assume that the group $G$ acts on $\MM_{\Scal}$ through some (again appropriately continuous) group action of automorphisms $\alpha:G\to \Aut(\MM_{\Scal})$. 
We may then define the \emph{relativisation map} $\yen:\MM_{\Scal}\to (\MM_{\Scal}\otimes B(\HH_{\Rcal}))^G$ into the $G$-invariant joint system-reference algebra, by
\begin{equation}
\yen(A)=\int_{G} \alpha(g,A)\otimes \diff E(g).
\end{equation}

In general, the map $\yen:\MM_{\Scal}\to (\MM_{\Scal}\otimes B(\HH_{\Rcal}))^G$ is not surjective. The full algebra of invariants can be described as follows.
\begin{theorem}
	There exists a Hilbert space $\KK$ and a unitary map $V:\HH_{\Rcal}\to L^2(G)\otimes \KK$ such that
	\begin{equation}
	(\MM_{\Scal}\otimes B(\HH_{\Rcal}))^G=(\id_{\HH_{\Scal}}\otimes V)^*((\MM_{\Scal}\otimes B(L^2(G)))^G\otimes B(\KK))(\id_{\HH_{\Scal}}\otimes V),
	\end{equation}
	where $L^2(G)$ is defined w.r.t. the left Haar measure and $G$ acts on $L^2(G)$ via the left action $(g.\psi)(h)=\psi(g^{-1}h)$ for $g,h\in G$ and $\psi\in L^2(G)$. 
\end{theorem}
This is a special case of \cite[Thm.~4.9]{fewsterQuantumReferenceFrames2025}. Here the space $\KK$ and map $V$ can be constructed explicitly from the QRF using a group representation theoretic result known as Mackey's imprimitivity theorem \cite{mackeyUnitaryRepresentationsGroup1958,cattaneoMackeysImprimitivityTheorem1979}.
The algebra $(\MM_{\Scal}\otimes B(L^2(G)))^G$ is a well-understood object in the theory of von Neumann algebras and goes under the name \emph{crossed product algebra} $\MM_{\Scal}\rtimes_{\alpha}G$ \cite{vanDaele:1978}. Crossed product algebras have become a recent focus in theoretical physics due to their appearance in models for quantum gravity, where they play a role in defining a notion of entropy \cite{wittenGravityCrossedProduct2022a,chandrasekaranAlgebraObservablesSitter2023}. Below, we describe a generalisation of these results in the context of QFT on curved spacetimes combined with QRFs.
\section{Invariant algebras and type reduction for QFT on curved spacetimes}
We now consider a potentially curved globally hyperbolic spacetime $M$. Suppose this spacetime has a group of spacetime isometries $G$ leaving the physics invariant. In analogy with our example of special relativity, we take physical observables to be $G$-invariant, which may be defined relative to a (quantum) reference frame. This should in particular hold for $G$-covariant quantum field theories defined on $M$. In the conventional description of AQFT \cite{haagLocalQuantumPhysics1996}, the observables of a quantum field on a fixed spacetime $M$ form an algebra $\Af(M)$, and to each region $U\subset M$ (say, some open set), the observables measurable in $U$ form an algebra $\Af(M;U)$.\footnote{See e.g.~\cite{fewsterAsymptoticMeasurementSchemes2023} where the notion of measurability in some region is made more precise.} We assume here that these algebras are von Neumann algebras. We then take $G$-covariance of the QFT to mean that there is an (appropriately continuous) group action $\alpha:G\mapsto\Af(M)$ such $\alpha(g)(\Af(M;U))=\Af(M;g.U)$ for any spacetime isometry $g\in G$. 
If a classical reference frame is available to distinguish between each two regions $U$ and $g.U$, or any two observables $A\in \Af(M)$ and $\alpha(g)(A)$, it indeed makes sense to think of the full algebra $\Af(M)$ as observable. Instead, we replace this classical reference frame by a quantum reference frame $(\HH_{\Rcal},U_{\Rcal},E)$ for $G$. Following our previous discussion, we take physical observables of the joint QFT/reference system to be elements of the algebra $(\Af(M)\otimes B(\HH_{\Rcal}))^G$, i.e.~invariants under the action $\alpha\otimes \Ad U_{\Rcal}$.\footnote{Here $\Ad U_{\Rcal}$ denotes the adjoint action of $U_{\Rcal}:G\to B(\HH_{\Rcal})$ on $B(\HH_{\Rcal})$, i.e. $Ad U_{\Rcal}(g)(A)=U_{\Rcal}(g)AU_{\Rcal}(g^{-1})$ for $g\in G$ and $A\in B(\HH_{\Rcal})$.}

Unlike to what one is used to in the context of quantum mechanics, the von Neumann algebras $\Af(M;U)$ typically appearing in QFT are \emph{purely infinite}, meaning that for any well-behaved trace,\footnote{A trace on a ${}^*$-algebra $\mathcal{A}$ is a positive linear map on non-negative elements $\tau:\mathcal{A}^+\to [0,\infty]$ satisfying $\tau(x^*x)=\tau(xx^*)$.} the only trace-class operator in $\Af(M;U)$ is $0$.\footnote{More specifically, under very general assumptions on the QFT and the region $U\subset M$, the local algebras of $\Af(M;U)$ are type $\mathrm{III}_1$ factors \cite{buchholzUniversalStructureLocal1987}.} For more details on traces and von Neumann algebras, see e.g.~\cite[Apx.~D]{fewsterQuantumReferenceFrames2025}. As a consequence, notions of entanglement entropy in QM, which rely on traces through the von Neumann entropy formula \cite{neumannThermodynamikQuantenmechanischerGesamtheiten1927}, do not yield well-behaved results in QFT.\footnote{Informally, the entanglement entropy of the QFT between the region $U$ and its causal complement diverges.} In very special cases however, one finds that the algebra $(\Af(M)\otimes B(\HH_{\Rcal}))^G$ (and its relevant subalgebras) may be much better behaved on this front than $\Af(M)$, i.e.~ its trace class operators form a dense subset (\emph{semi-finiteness}) or the algebra may even admit a tracial state (\emph{finiteness}). Similarly to as discussed in \cite{chandrasekaranAlgebraObservablesSitter2023}, the mathematical understanding of this phenomenon, which in \cite{fewsterQuantumReferenceFrames2025} is referred to as \emph{type reduction}, relies on the interplay of crossed product algebras with Tomita-Takesaki modular theory \cite{takesakiDualityCrossedProducts1973}. In more physical terms, we can ascribe this phenomenon to thermodynamic properties of the QFT.

For simplicity, we assume the group $G$ on $M$ to be given by time translations, particularly $G\cong \RR$ (the set-up of \cite[Sec.~5]{fewsterQuantumReferenceFrames2025} also allows for larger groups). We assume that the von Neumann algebra $\Af(M)$ admits a well-behaved thermal state $\omega$.\footnote{A state is a positive normalised linear functional $\omega:\Af(M)\to \RR$. Thermality here refers to that $\omega$ satisfies the $\beta$-KMS condition at inverse temperature $\beta$ \cite{cmp/1103840050} with respect to the time translations $\alpha:\RR\to \Aut(\Af(M))$. We shall furthermore assume this state to be normal (i.e.~appropriately continuous) and faithful (i.e.~non-vanishing on positive operators).} Similarly, one can consider thermodynamic properties of the QRF. While the algebra $B(\HH_{\Rcal})$ will not admit any thermal states with respect to $\Ad U_{\Rcal}$,\footnote{This is a consequence of the spectral properties of $U_{\Rcal}$.} we can formulate a weaker thermality condition \cite[Eq.~5.17]{fewsterQuantumReferenceFrames2025}, which we shall refer to as \emph{local $\beta$-finiteness}. The following result is then a special case of \cite[Thm.~5.5]{fewsterQuantumReferenceFrames2025}.
\begin{theorem}
If $\Af(M)$ admits a well-behaved thermal state at inverse temperature $\beta$, then the algebra $(\Af(M)\otimes B(\HH_{\Rcal}))^\RR$ is semi-finite. If the QRF $(\HH_{\Rcal},U_{\Rcal},E)$ is furthermore locally $\beta$-finite, then the algebra $(\Af(M)\otimes B(\HH_{\Rcal}))^\RR$ is finite.
\end{theorem}
We may hence say that, if both the QFT and QRF have `good thermal properties' at the same temperature, then their joint algebra of invariants admit a finite trace. Consequently, it is thus in particular also finite on all subalgebras of $(\Af(M)\otimes B(\HH_{\Rcal}))^\RR$. The algebras discussed in \cite{wittenGravityCrossedProduct2022a,chandrasekaranAlgebraObservablesSitter2023} and subsequent works can all be realised as examples of the algebras addressed by Thm.~2. In particular, the models discussed in these works all give rise (or are appended with) an operational quantum reference.\footnote{For similar observations through another approach to QRFs, see \cite{vuystCrossedProductsQuantum2025}.}

\section{Gauge theories, boundaries and quantum reference frames}
In various models of QG, the physical degrees of freedom that give rise to time-translation QRFs are, in one way or another, associated with a spacetime (or region) boundary \cite{chandrasekaranLargeAlgebrasGeneralized2023,kudler-flamGeneralizedBlackHole2025}. More generally, it has been observed that when one considers a gauge theory on a spacetime with boundaries, one may associate degrees with the boundary that give rise to reference frames for gauge transformations at the boundary \cite{donnellyLocalSubsystemsGauge2016,carrozzaEdgeModesReference2022,kabelQuantumReferenceFrames2023}.
In \cite{fewsterSemilocalObservablesEdge2025}, this is made precise using an algebraic formulation of quantum electromagnetism on a class of spacetimes with boundaries and corners, dubbed \emph{finite Cauchy lenses}. 

At each Cauchy surface with boundaries $\ol{\Sigma}$, the observables of the theory can be expressed in terms of smearings of the EM vector potential $\mathbf{A}$ and electric field $\mathbf{E}$ (understood as differential one-forms on $\ol{\Sigma}$). These smearings are such that the observables are invariant under gauge transformations $\mathbf{A}\mapsto \mathbf{A}+\diff \Lambda$, where $\Lambda$ is a smooth function on $\ol{\Sigma}$ vanishing on its boundary $\partial\ol{\Sigma}$. 
\begin{wrapfigure}{r}{7.5cm}
\includegraphics[width=7.5 cm]{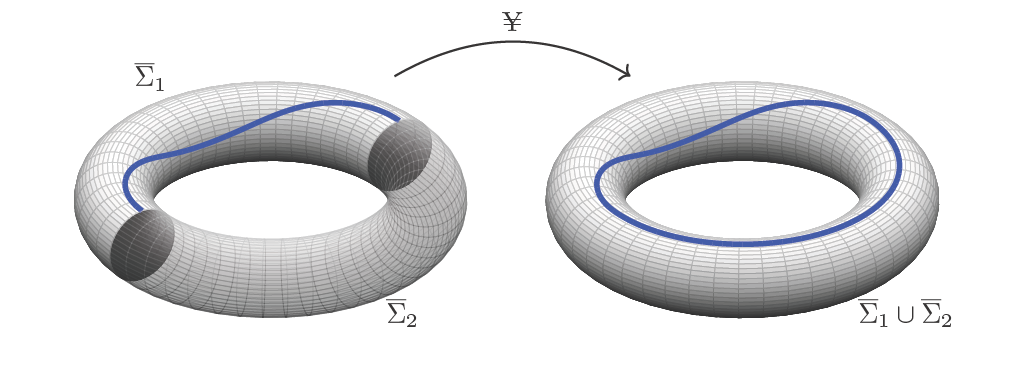}
\caption{A Cauchy surface $\ol{\Sigma}$ is divided in $\ol{\Sigma}_1$ and $\ol{\Sigma}_2$. By treating the electromagnetic field on $\ol{\Sigma}_2$ as a QRF for $\ol{\Sigma}_1$, one can build large gauge invariant observables on $\ol{\Sigma}$ from edge mode observables through relativisation \cite[Fig.~4]{fewsterSemilocalObservablesEdge2025}.}\label{fig:glue}
\end{wrapfigure} 
This notion of gauge invariance is sufficient to ensure well-behaved dynamics of the theory. The resulting algebra contains observables that transform covariantly under \emph{boundary/large gauge transformations} (i.e.~for which $\Lambda$ does not vanish on $\partial\ol{\Sigma}$). These are dubbed \emph{edge-mode observables} and can be understood as (smeared analogues of) Wilson lines connecting two points on the boundary $\int_0^1\mathbf{A}_\mu \diff\gamma^\mu$ for a curve $\gamma:[0,1]\to \ol{\Sigma}$ with $\gamma(0),\gamma(1)\in\partial\ol{\Sigma}$. By quantising the Poisson structure arising naturally from the \emph{covariant phase space formalism} for electromagnetism on spacetimes with boundaries \cite{harlowCovariantPhaseSpace2020}, these edge mode observables have non-trivial commutation relations with smearings of the electric flux density $\nml_{\partial\ol{\Sigma}}\mathbf{E}$ through $\partial\ol{\Sigma}$. As a consequence, these fluxes are quantised, rather than superselected as in conventional approaches to quantum electromagnetism \cite{buchholzPhysicalStateSpace1982}. 

At the level of sufficiently regular Hilbert space representations, the edge mode observables can be used to construct an operational quantum reference frame for the boundary gauge group, see \cite[Sec.~5]{fewsterSemilocalObservablesEdge2025}. The associated relativisation map can in particular be used to glue together algebras of observables associated with neighbouring regions, as well as their states, see \cite[Sec.~6]{fewsterSemilocalObservablesEdge2025}. This is illustrated in Fig.~\ref{fig:glue}.

\section*{Conclusions}
Quantum reference frames offer a way of constructing algebras of invariants with a rich structure in quantum theories. Their role in type reduction phenomena in QFT and QG suggest they might play a central role in defining entropy in quantum gravity. More generally, their presence impacts the quantisation of gauge theories on spacetimes with boundaries. The combination of algebraic QFT as well as the operational QRFs provides a rigorous mathematical setting where these phenomena can be analysed.

\paragraph{Acknowledgements} This work was supported by EPSRC Grant EP/Y000099/1 to the University of York.
For the purpose of open access, the authors have applied a creative commons attribution (CC BY) licence to any author accepted manuscript version arising.

\bibliographystyle{iopart-num}
\bibliography{Refs}

\end{document}